# USING FISHER INFORMATION IN BIG DATA

Nasir Ahmad[a,c], Sybil Derrible[a], Tarsha Eason[b] and Heriberto Cabezas[b]

[a]Complex and Sustainable Urban Networks (CSUN) Laboratory, University of Illinois at Chicago, Chicago, IL [b]Sustainable Technology Division, Office of Research and Development, U.S. Environmental Protection Agency, Cincinnati, OH
[c](corresponding author) sahmad38@uic.edu

Abstract – In this era of Big Data, proficient use of data mining is the key to capture useful information from any dataset. As numerous data mining techniques make use of information theory concepts, in this paper, we discuss how Fisher information (FI) can be applied to analyze patterns in Big Data. The main advantage of FI is its ability to combine multiple variables together to inform us on the overall trends and stability of a system. It can therefore detect whether a system is losing dynamic order and stability, which may serve as a signal of an impending regime shift. In this work, we first provide a brief overview of Fisher information theory, followed by a simple step-by-step numerical example on how to compute FI. Finally, as a numerical demonstration, we calculate the evolution of FI for GDP per capita (current US Dollar) and total population of the USA from 1960 to 2013.

Keywords—*Fisher information*, *Data analysis*, *Big Data*

## I. INTRODUCTION

As we pass through the 21st century, a massive advancement of information technology is experienced by almost all the sectors around us. There is a long history of manipulating information by humans but in modern days the process has become more advanced and robust. Efforts have been made to gain more information not only about observable phenomena but also about latent parameters inherent in system data. Data mining and Big Data have facilitated the growth and development of a plethora of complex approaches to manipulate and capture patterns in system data. Many of these concepts originate from information theory. Rooted in statistics, information theory resides between computer science, mathematics, physics and engineering and has been widely applied from cryptology to ecosystem dynamics [1]. Fisher information, a key method in information theory, offers great promise for data mining applications. It was developed by Ronald Fisher [2] as a means of measuring the amount of information about an unknown parameter that can be obtained by observations. Since then, it has been adapted into a means of monitoring system variables to assess patterns and evaluate stability in system dynamics [3].

Fisher information has been used in a variety of applications from deriving fundamental laws of thermodynamics [4] to assessing dynamic order in real and model systems [3], [5–8] to sustainable environmental management and resilience [9–16]. However, there has been limited use in the engineering arena.

Based on previously developed methods, this paper presents a brief overview of Fisher information theory, the calculation algorithm and provides a simple computational example. To demonstrate the utility of Fisher information for data mining applications, the approach is used to evaluate the evolution of GDP per capita (current US$) and total population of the USA from 1960 to 2013. While methods for computing Fisher information have been developed in Matlab [10], we have translated the Matlab code into an open source Python library that can be freely downloaded from GitHub (https://github.com/csunlab/fisher-information).

## II. BACKGROUND ON FISHER INFORMATION

### A. Definition of Fisher Information

Fisher information (FI) was first developed by statistician R.A Fisher [2], and it offers a measure of indeterminacy. In other words, FI can measure the amount of information of an unknown parameter that is present in observable data. Mathematically, the Fisher Information (FI), *I(θ)*, is defined as [3]:

$$I(\theta) = \int \frac{dX}{p_0(X \mid \theta)} \left[ \frac{\partial p_0(X \mid \theta)}{\partial \theta} \right]^2 \quad (1)$$

where, *p₀(X|θ)* is the probability density of obtaining a particular value of *X* in the presence of *θ*.

In practice, it is essentially impossible to use equation 1 because the computation of the derivative of the ($\partial p_0(X|\theta) / \partial \theta$) component is required for this process, which depends on the numeric value of the unknown parameter $\theta$. Through numerous derivation steps, Mayer et al. [8] adapted this equation for application to real systems:

$$I = \int \frac{ds}{p(s)}\left[\frac{dp(s)}{ds}\right]^2 \quad (2)$$

Based on the probability of observing various states of the system $p(s)$. Equation 2 is the foundational form of Fisher information used in this work. The equation is further simplified by transformation to eliminate the challenge of a small $p(s)$ in the denominator. To overcome this problem, akin to quantum mechanics, we replace $p(s)$ by its magnitude $q^2$, which after some manipulation gives [8]:

$$I = 4\int \left[\frac{dq(s)}{ds}\right]^2 ds \quad (3)$$

Karunanithi et al. [3] further simplified this equation by assuming discrete steps so that $dq \approx \Delta q = q_i - q_{i+1}$ and $ds \approx \Delta s = s_i - s_{i+1}$. For consistent time steps, $s_i - s_{i+1} = 1$. Therefore, equation 3 can be written as:

$$FI \approx 4\sum_{i=1}^{m}[q_i - q_{i+1}]^2 \quad (4)$$

where, $m$ is the number of states. A state is defined as a condition of the system determined by specifying a value for each of the variables that characterize its behavior [3]. Equation 4 can now be used to compute Fisher information numerically for systems characterized by discrete data. The following section will discuss the step-by-step procedure to compute Fisher information (henceforth denoted as FI). Complete details on this and related derivations may be found in [3], [5], [9], [13].

## III CALCULATION METHODOLOGY

Evaluating changes in the probability of detecting different states of a system over time is the foundation of the computation of FI. Hence, information about a system's condition or state over time is required. A system can be defined by $n$ measurable variables ($y_i$), which are able to characterize the system and its state at any point in time [3]. Each data point $v_i$ at time $t_j$, ($v_{i,j}$), in a phase space is defined by the set of variables $v_{i,j} = \{y_1(t_j), y_2(t_j), \ldots, y_n(t_j)\}$, when the system can easily be categorized in discrete states. In practice, small fluctuations in a variable do not systematically translate into a phase change. Moreover, some inherent or measurement error also frequently occur. We define these fluctuations and small errors as uncertainty in our system.

Estimation of the uncertainty is critical to define the states of the system. Numerically, a parameter $\Delta y_i$ is defined as measurement uncertainty such that, if:

$$|y_i(t_j) - y_i(t_k)| \leq \Delta y_i \quad (5)$$

is true for all variables $y_i$ at time $t_j$ and $t_k$ then the two points are in the same phase and consequently "binned" together in the same state. In other words, if a system is defined by $n$ measurable variables then a state is exemplified as $n$ dimensional hyper-rectangular box, where each side represents the uncertainty for each variable. Here, this set of $\Delta y_i$ defines the size of state for the system.

Usually, unless reported with the data, the measurement of uncertainty is unknown. Karunanithi et al.[3] recommend choosing a relatively stable time period in each time series, and then computing the standard deviation (SD) of each variable $Y$ with population mean $\vartheta$ and using Chebyshev's inequality, defined by:

$$P(|Y - \vartheta| < k \cdot SD) \geq (1 - \frac{1}{k^2}) \quad (6)$$

Equation 6 indicates that for any form of probability distribution, "the proportion of the observations falling within $k$ standard deviations of the [population] mean ($\vartheta$) is at least $1-1/k^2$"[17]. Thus, $\Delta y_i$ is chosen as $\pm k \cdot SD$. To ensure at least 75% of the data would occur within the level of uncertainty a $k$ of 2 can be selected as $1-1/2^2=0.75$.

In other words, for one variable, two points can be considered to belong to the same phase if they vary within a defined level of uncertainty for this variable. Overall, this means that the state of a system is represented by all the points that are "binned" within a range of uncertainty rather than one single point in time [3].

As mentioned earlier, the goal of FI is to capture the dynamic behavior in terms of the probability of observing various states of a system. To move through the data, the time period is divided into time windows composed of several time steps (e.g., eight consecutive years), and one measure of FI is calculated for the time window which we attribute to the last time step of the window so that only past data are used in the computation. The time window is then moved by a defined number of steps. Two parameters are therefore used to define the moving window, which are the size of the window and the increment of the window. Both of these parameters are expressed in terms of time steps. These two parameters are used to move through the data such that the size of the window is greater than the amount of movement for each window. The numerical example below illustrates this point. Then, the probability densities $p(s)$ and eventually FI for each window are computed. The size of the window depends on the size of the data but it has been found empirically that the window size should be at least eight time steps. Further details on the computation algorithm can be found in the USEPA report published in 2010 [10].

After determining the parameters for the integration window (window size, window increment and size of state), the binning process can be initiated. To begin, the first point of the time window is selected as the center of the first state and a hyper-rectangle, whose sides are defined by $\Delta y_i$ for each variable of that system, is placed around that point. The points that lie within the hyper-rectangle are binned together. Then, the next unbinned point in the window is taken as the center of the hyper-rectangle and similar points found within that hyper-rectangle are binned together. This process continues until all the points in the time window are binned or placed in different states.

Table 1 Sample Data for a Time Window

| Time Step | $Y_1$ | $Y_2$ |
|---|---|---|
| 1 | 0.6 | 1.5 |
| 2 | 2 | 1.5 |
| 3 | 0.3 | 1 |
| 4 | 3.5 | 4.8 |
| 5 | 0.95 | 2 |
| 6 | 3.1 | 4 |
| 7 | 2.4 | 1.8 |
| 8 | 2.7 | 2.1 |
| $\Delta Y$ | 0.5 | 1 |

Following the approach presented by USEPA [10], as a small numerical example, Fig. 1 shows a binning process for a two dimensional system, which is defined by two variables with size of state of 0.5 and 1 consecutively (Data shown in Table 1). Eight time steps are defined in each window and result in one measure of FI, which is plotted at the end of the window. For example, time steps 1 to 8 could represent data from year 2001 to 2008. For this example, we assign the value of FI to time step 8 (e.g.,

2008). The next time window will go from time 2 to 9 (e.g., 2002 to 2009), followed by time 3 to 10, etc.

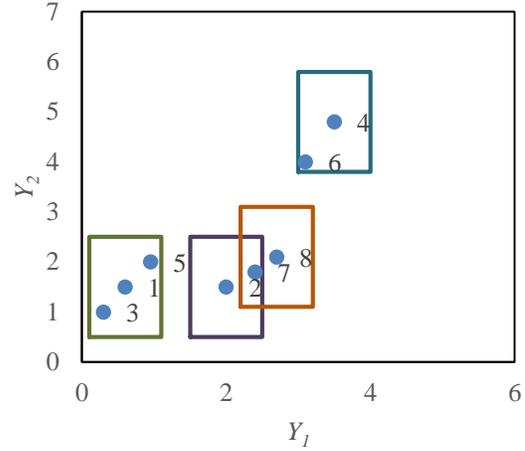

Figure 1 Binning Process

When, all the points are binned together, then probability distribution ($P_i$) for each window is measured by using equation 7 [10]:

$$Pi = \frac{\text{Number of points in state}}{\text{Total number of point is window}} \quad (7)$$

The probability distribution for the sample data in Table 1 is shown in Fig. 2

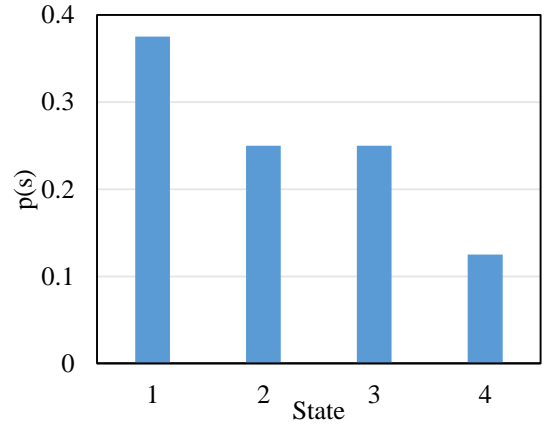

Figure 2 Probability Distribution

Then the amplitude, $q$ ($q_i = \sqrt{p_i}$) and FI for each window is calculated by using equation 4, where the initial and final $q_i$ is set as zero. Figures 2 and 3 display the $p(s)$ and $q(s)$ for each state based on the sample data in Table 1. The FI for the sample data using Equation 10 is: 4x(0.375+0.13+0+.021+.125) = 4x0.534 = 2.136

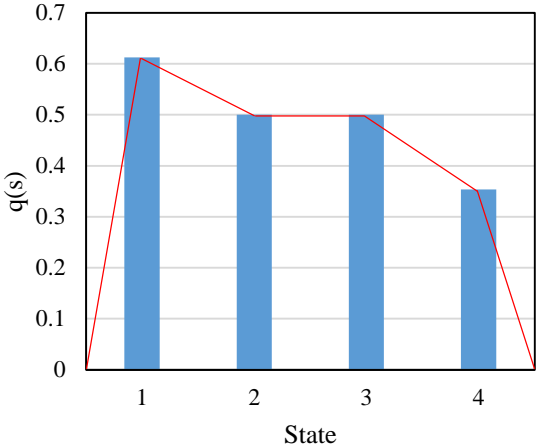

Figure 3 Amplitude of the Probability Distribution

## IV INTERPRETATION OF FI

The Sustainable Regimes Hypothesis was developed to provide a construct for interpreting FI [3], [5], [9].

- A system is considered to be in an orderly dynamic regime when a non-zero FI remains nearly constant over time (i.e., d<FI>/d$t \approx 0$).
- A steady decrease in FI insinuates that the system is losing its functionality, stability and the patterns are breaking down. This declining trend may provide warning of an impending regime shift [15]
- A steady increase of FI indicates that the system is becoming more organized / stable.

When using the velocity and acceleration approach to compute FI, it is possible to observe peaks in FI over time [11]. These represent not points of stability but inflection points, i.e., points in time when the system trajectory in its phase space changed direction. This is important because repeated changes in direction signify a system that is unable to find a stable regime path through time.

Researchers have studied the behavior of Fisher information in the neighborhood of a tipping point [18]. Whereas most systems tend to exhibit declining FI as a warning of impending transitions [15, 16], a number of theoretical scenarios have been explored to model expected behavior under different conditions [18]. From this study, it is clear that the behavior of Fisher information depends heavily on the trends in the variables as the system approaches a tipping point.

## V NUMERICAL DEMONSTRATION

To demonstrate the procedure on a real dataset, we analyzed the evolution of GDP per capita (current US$) and U.S. population from 1960 to 2013. The data were collected from the World Bank data catalogue [19]. Of course, these two variables may not be enough to capture how the economy performed from 1960 to 2013, but the example demonstrates how FI could inform us about patterns and stability in the system, which is defined by the two variables. The window size of 8 and window increment of 1 is chosen for this example. The size of state is calculated by using the algorithm mentioned earlier and found to be 985.82 and 10,307,105.62 respectively for the two variables. FI is then calculated and depicted in Fig. 4. Because we assign the FI value to the end of the window, the first value reported is for 1967, representing 1960 to 1967. Hence, the figure shows data for years 1967 to 2013.

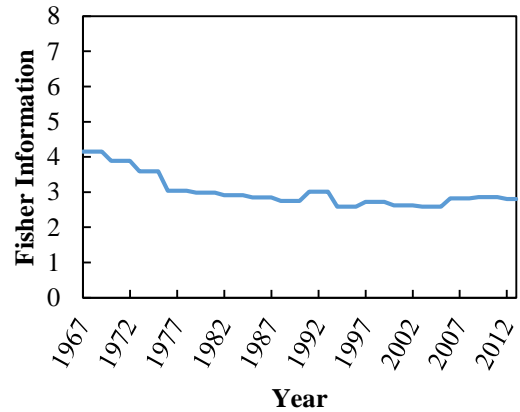

Figure 4 Fisher Information for GDP per capita and total population of the USA from 1967 to 2013

Except for an initial decrease, Fig. 5 shows that FI remain relatively stable from 1967 to 2013. According to the Sustainable Regimes Hypothesis mentioned in section IV, this observation suggests that the system is considered well-functioning with an orderly dynamic regime. The selection of variables is critical to characterizing system behavior and needs to be carefully considering for FI to output meaningful results.

## VI CONCLUSION

The main objective of this paper was to introduce Fisher information as a useful method for assessing patterns in Big Data. We first provided a brief summary of the theory and algorithm used to compute FI. We then used a small numerical example to illustrate how FI is computed. Finally, we calculated

FI to measure the stability of GDP and population in the United States from 1960 to 2013. The key strength of FI is its ability to handle systems defined by multiple variables. Overall, these properties make FI a viable tool to assess patterns in complex systems and it therefore possesses much utility for data mining.

**Acknowledgement** — This research was supported, in part, by NSF Award CCF-1331800. Any use of trade, firm, or product names is for descriptive purposes only and does not imply endorsement by the U.S. Government. The views expressed are those of the authors and do not necessarily represent the views or policies of the U.S. Environmental Protection Agency.